\documentclass[10pt]{sig-alternate-10pt}   	
\usepackage{authblk}
\usepackage{graphicx}				
\usepackage{amssymb}
\usepackage{caption}
\usepackage{subcaption}
\usepackage{url}

\title{An Evaluation of Yelp Dataset}
\author{YAN CUI}
\affil{Columbia University}

\begin{document}
\maketitle
\section{Abstract}
Yelp is one of the largest online searching and reviewing systems for kinds of businesses, including restaurants, shopping, 
home services et al.  Analyzing the real world data from Yelp is valuable in acquiring the interests of users, which helps to 
improve the design of the next generation system. This paper targets the evaluation of Yelp dataset, which is provided in
the Yelp data challenge. A bunch of interesting results are found. For instance, to reach any one in the Yelp social network, 
one only needs 4.5 hops on average, which verifies the classical six degree separation theory; 
Elite user mechanism is especially effective in maintaining the healthy of the whole network; Users who write less than 
100 business reviews dominate. Those insights are expected to be considered by Yelp to make intelligent 
business decisions in the future. 
   
\section{Dataset}
The dataset is collected from real Yelp business and is released for academic purpose~\cite{dataset}.  It contains 366K user 
profiles for a total of 2.9M edges; 1.6M reviews and 500K tips for 61K businesses and aggregated check-ins over time for 
each of the 61K businesses. To make the dataset diversity, the data of four countries are included (U.S., UK, Germany and 
Canada). All the data are divided into 5 JSON files, which are business, review, user, checkin and tip. The content of these files 
are straightforward to understand. Specifically, business file presents the attributes of each business, like the internal unique id of the 
business, name, geolocation, categories, et al. The checkin file is a complementary attributes for each business. It presents
the aggregated checkin time for each business in the grain of each hour in a week. User file contains the basic information for 
each user, such as yelping time, name, fans and social relationship with other users. Tip and review files are the review of a 
particular user to a business. The difference of tip and review is that tip is always short to represent a user's opinion, like 
"Don't waste your time." or "Great drink specials!", while review is long enough to allow users express their feedback.

\section{Data Storage}
Due to the size of dataset, it is both time consuming and inconvenient to analyze the statistics without the help of data storage 
system. At the same time, it is rather natural to model the network of Yelp using graphs, because users and businesses can be
abstracted as nodes with different properties; review to a business and user friendships can be modeled as the relationships 
between nodes. Thus, the graph database is a great fit for this purpose. For this project, we select Neo4j~\cite{neo4j} as the
database as it has a great web UI which makes graph visualization possible and a SQL-like query language (Cypher), which is easy 
to learn and use. In our graph, there are two kinds of nodes, \textit{User} and \textit{Business}. 
The information used to describe users and businesses are stored as attributes in the node. Three relations exist in 
the graph, which are \textit{knows}, used for user friend relationship, \textit{tip} and \textit{review} for user and business 
relationships, respectively. 

Importing the dataset into Neo4j is all about creating nodes or relationships. However, it is worth to mention two tricks we 
made. The first one is used when importing user friendship data into Neo4j. Each user has zero to many friends. To build up 
the friendship, what we can do is a straightforward twice scan algorithm. We create each user node without adding the 
friendship at the first scan, and build up relationship at the second time. To improve the performance of data importing, we 
adopt a scan-once method. The basic idea is when creating a user node, we build up the friend relationship directly. 
Specifically, for each friend of a user, a look up is performed in the current graph. 
If a node is found, the user node being established and the found node is connected using \textit{knows} relationship, 
otherwise, we create a node with label \textit{Person}, which means it will become a user in future, but right now its full 
set of properties are not filled.  When trying to create a new user, a look up is also necessary. If there is not such a node, 
we create a new user, otherwise, the user node exists as a \textit{Person}, and the full information should be filled into the node. 
exist as a \textit{Person} node. Another trick that we used to speedup the dataset importing is to create unique index 
for both users and businesses because we noticed that the internal user id and business id are unique and there are 
a lot of queries made based on their ids.
  
Although the use of these tricks, the whole importing process is also quite slow. Roughly speaking, it costs us 2 weeks to 
import all 366K user basic information and their relationships. Although one can optimize the graph establish using REST API 
instead of Cypher, it is left as future work. In order to establish graph quickly and reproduce our experiments results easily, 
our database is open sourced using Dropbox, along with the graph establish and analysis codes used in this paper (https://github.com/ycui1984/yelp-data-challenge). Readers who are interested to use our database just need to download the database, extract
in the proper folder and restart the service. 

\section{Analysis}
The benefits of using Neo4j is that it simplifies our analysis process by writing simple Cypher queries. For complex 
tasks, we created python scripts and used \textit{py2neo} library to manipulate Neo4j. 

\begin{figure}[t]
  \centering
  \caption{Number of Yelp Users against Degree} 
       \includegraphics[width=0.48\textwidth]{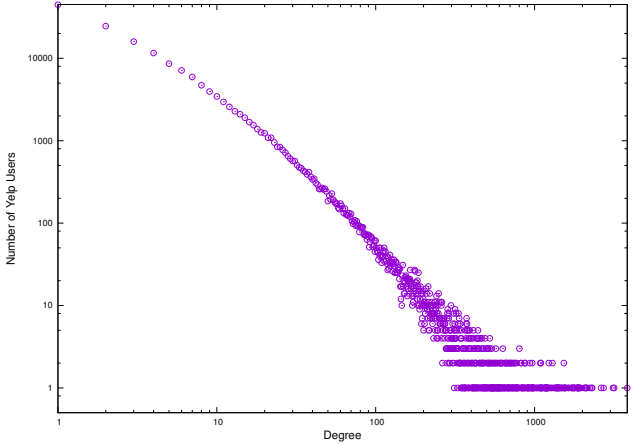}
  \label{fig:degree-dis}
\end{figure}

\subsection{Number of Friends Distribution} 
From the high level, we found that there are 2576179 \textit{knows} directed friend relationships among 366715 users. 
So, on average, each user has roughly 7 friends. To investigate into the number of friends distribution for each user, 
we looked at the number of Yelp users for different degrees, which is shown in Figure~\ref{fig:degree-dis}. In this figure, 
we ignore 192621 users with no friends. One thing worth to point out is that x and y axises use the log scale. 
As shown in this figure,  the degree distribution exhibits the power law phenomenon, which is expected. 
In the dataset, the largest degree among all users is 3830.  Although there is a hard limit for the largest number of 
friends for a user on Yelp (5000)~\cite{limit}, we did not see any user who hits the limit.   

\begin{figure}[t]
  \centering
  \caption{CDF for Number of Hops between Two Nodes} 
       \includegraphics[width=0.48\textwidth]{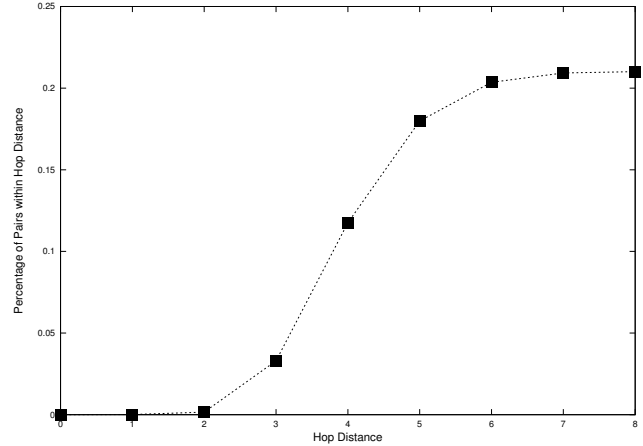}
  \label{fig:hops}
\end{figure}

\subsection{Hop Length between Users}
The number of hops between users are investigated in this subsection. A classical conclusion about social network is
\textit{six degrees of separation}~\cite{sixdegree}, which means that the average number of hops between every two nodes 
is no more than 6. To verify this, we randomly selected 10000 node pairs among all nodes and measured length of shortest
path. Based on the measurement, the cumulative distribution function (CDF) against the number of hops is calculated. 
Figure~\ref{fig:hops} 
presents the experimental results. As we can see, the CDF curve increases with the number of hops, and stabilizes 
at 8 hops. The probability of two users within 8 hops is 21\%. Actually, most of the time (79\%), there is no path 
between selected user nodes. One natural reasoning is that two nodes are selected from two separate connected components 
when there is no path, otherwise, nodes come from the same connected component.  Also, there should be a lot of connected 
components such that there is no path between two nodes in a high probability. Based on these conclusions, 
we can see the largest number of  hops between any two nodes in the same connected component is 8, 
which is the diameter of the connected component. The average distance of two nodes in the same connected 
component is calculated to be 4.5, which verifies the six degree separation theory. Note that the average path 
length is even smaller than most of social networks including Facebook~\cite{facebook}, YoutTube~\cite{imc2007}, 
and LiveJournal~\cite{imc2007}, which implies that the information in Yelp network spreads faster than other networks.
   
\begin{figure}[t]
  \centering
  \caption{Connected Components Distribution} 
       \includegraphics[width=0.48\textwidth]{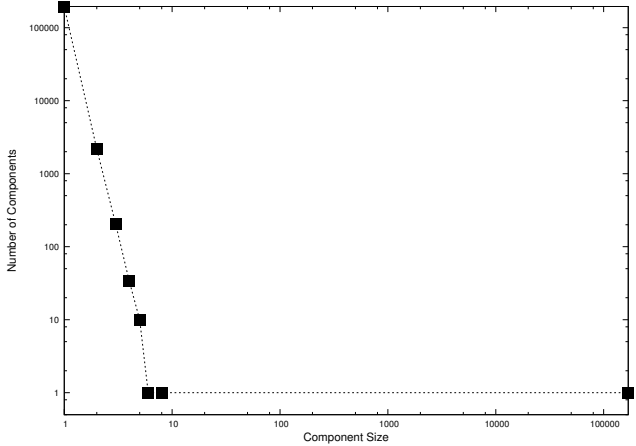}
  \label{fig:ccd}
\end{figure}   

\subsection{Connected Components Distribution}
To verify our reasoning about the structure of Yelp network, this subsection investigates the distribution of connected 
components. To find all components and the number of nodes in each component, we recorded each subgraph encountered 
so far.  Then, each user node is iterated one by one. For each subgraph, the path existence between the subgraph and 
the target node was tested. If there exists a path, the user node is included into the subgraph, otherwise, a new subgraph is
created. Figure~\ref{fig:ccd} presents the experimental results. As expected, there is 192621 connected components where
single user exists. Except that, most of users belongs to a single large connected component, which includes 168917 users. 
If we calculate the probability of two user nodes in the same connected component, it is roughly 0.21, which matches
the results in the last subsection.  

\begin{figure}[t]
  \centering
  \caption{Cluster Coefficient and Maximal K-Core} 
       \includegraphics[width=0.48\textwidth]{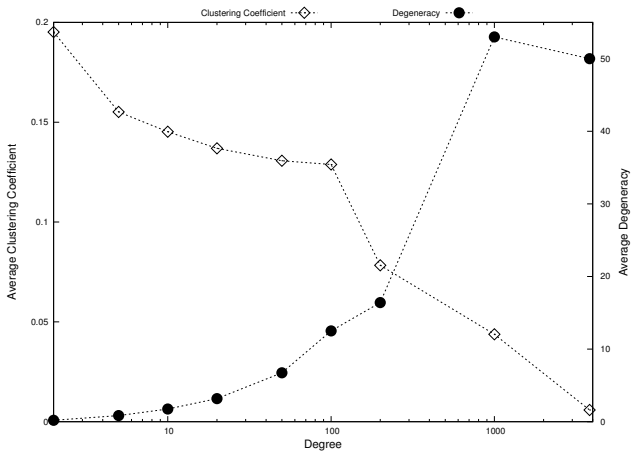}
  \label{fig:coefficient}
\end{figure}  

\subsection{Clustering Coefficient and Maximal K-Core}
After understanding the network structure in the high level, the next investigation is how Yelp friends of a particular user 
interact with each other. We try to answer interesting questions such as how many of your friends are also friends? To this 
end, we measure the average clustering coefficient for users with different degrees. The clustering coefficient for a 
graph is calculated by dividing the existing number of  \textit{knows} relationships by the total possible number 
of relationships. For example, assume there are  4 nodes in a graph, and there are  2 \textit{knows} relationships. 
So, the clustering coefficient is $2 / (4*3/2)$ = 0.33. Figure~\ref{fig:coefficient} (labeled as clustering coefficient) shows the results 
for users with different degrees.  As indicated, the coefficient keeps decreasing with the increasing degree, which 
suggests that the more friends one has, the more possible ways one's friends are also friends. Specifically, the total 
number of relationships is a number of $O(n*n)$, where n is the number of friends for a user, while the actual graph is 
not dense enough to keep up with the total number. To shed light on the structure of the subgraph formed by all friends of 
a user, we also measured the average maximal k-core for users with increasing degrees. The k-core of a graph is defined 
as the maximal subgraph where each node has k degrees at least. The maximal k-core is to look for a k-core 
which maximizes k.  The curve labeled as degeneracy in Figure~\ref{fig:coefficient} plots the experimental results. 
Overall, the trend is the more friends 
one has, the larger the maximal k-core. For instance, if one has 100 friends, the average maximal k-core is roughly 13,
which means there are at least 14 users who knows the other 13 friends of this user.  One exception of the curve is 
the data point when the degree is 1000. It is larger than that when degree is 3830. After examining the results, we found
that there is only one user who has degree 1000 and 3830. The exception is considered as statistical errors. If enough 
users with high degrees exist, the curve is believed to increase continuously. 

\begin{figure}[t]
  \centering
  \caption{Number of Friends of Friends} 
       \includegraphics[width=0.48\textwidth]{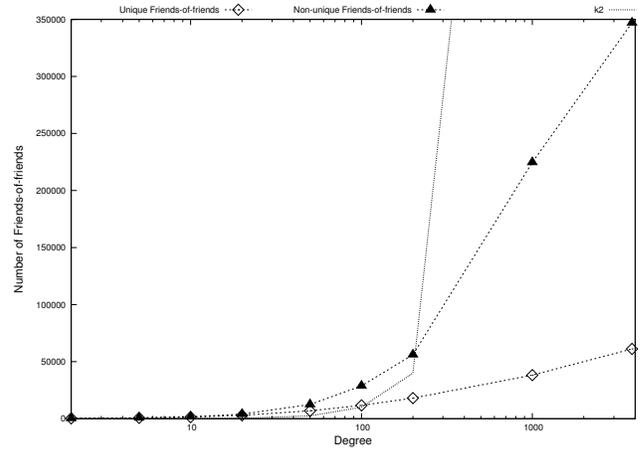}
  \label{fig:fof}
\end{figure}

\subsection{Friends of Friends}
One important property of social networks is friends of friends. For example, Linkedin recommends people you may know
if two users share the same friend. Yelp can also benefit from recommending to a user the businesses 
that liked by his friends of friends. To this end, the number of unique friends of friends and non-unique friends of friends with 
increasing number of degrees are counted. Note that unique friends of friends are a distinct set of user nodes one can 
reach by two hops from a user, while non-unique friends of friends are all user nodes that are reachable by two hops.  
Figure~\ref{fig:fof} plots the results. Besides, the $k*k$ curve is also presented for comparison, where $k$ is degree.  
The common sense is that if one has $k$ friends, then $k*k$ is an estimation of the number of total friends of friends, assuming
that each of his friends has $k$ friends. By adding the $k*k$ curve, one can compare with our estimation. From this 
figure, we can see that the number of unique friends of friends and non-unique friends of friends is larger than our 
estimation when the degree is smaller than 100 and 200, respectively. Note that the percentage of users whose 
degree is smaller than 100 is over 90\%, which suggests that, for most of users in Yelp network, your friends 
have more friends than you. 

\begin{figure}[t]
  \centering
  \caption{Average Neighbor's Degree} 
       \includegraphics[width=0.48\textwidth]{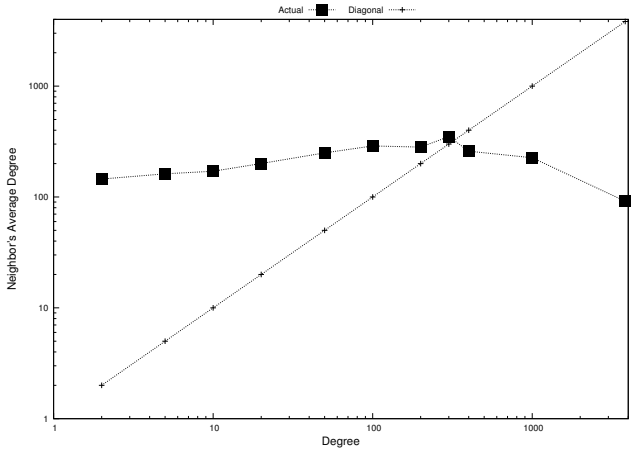}
  \label{fig:avg}
\end{figure} 

To further prove this conclusion, the average neighbor's degree against degree is presented in Figure~\ref{fig:avg}.
Note that the figure is drawn in log scale for both x and y axises. As can be seen, the average neighbor's degree 
keeps increasing when the degree of a user is relatively small, which verifies that your friends has more friends than
you. Once the degree is large enough, the average neighbor's degree becomes smaller. However, this only 
happens to a small portion of users. 

\begin{figure*}[t]
  \centering
  \begin{subfigure} [b]{0.45\textwidth}        
  	  \includegraphics[width=0.99\textwidth]{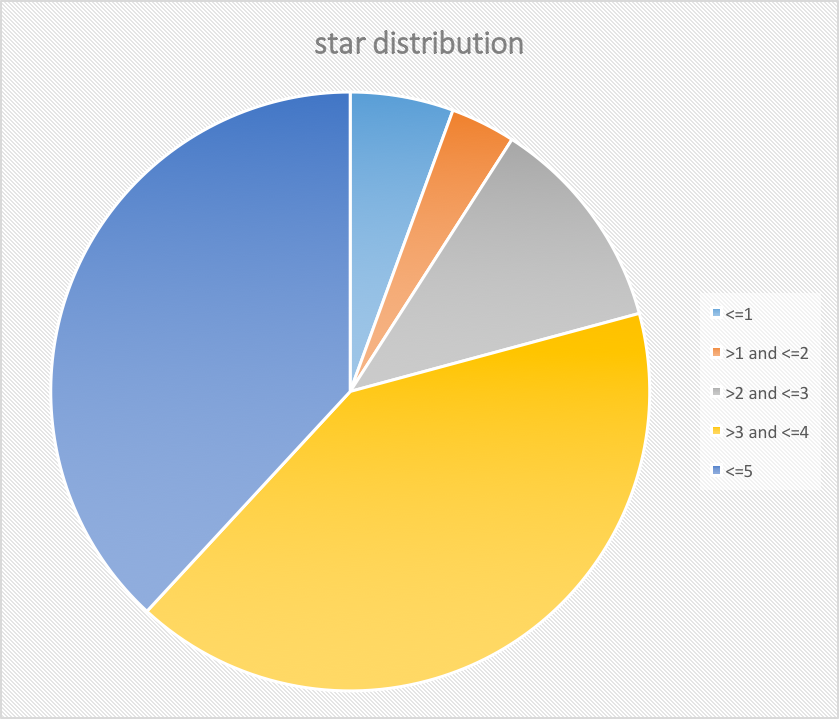}
        \caption{Star Distribution}
        \label{fig:star}
  \end{subfigure}
  \begin{subfigure}[b]{0.45\textwidth}        
        \includegraphics[width=0.99\textwidth]{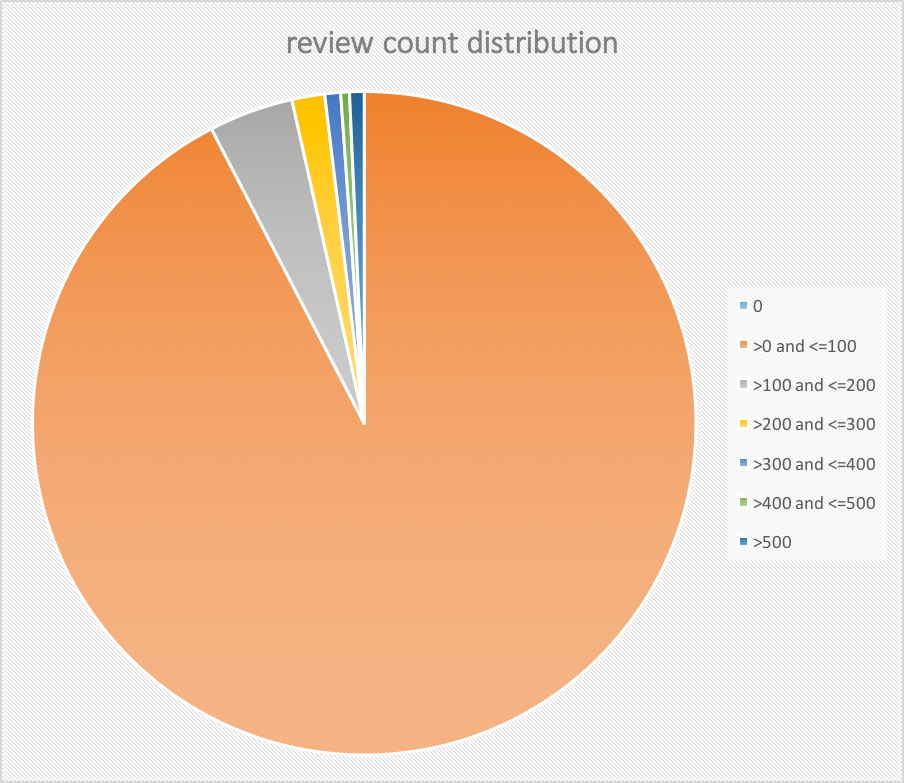}
        \caption{Review Distribution}
        \label{fig:review}
  \end{subfigure}
  \begin{subfigure}[b]{0.45\textwidth}        
        \includegraphics[width=0.99\textwidth]{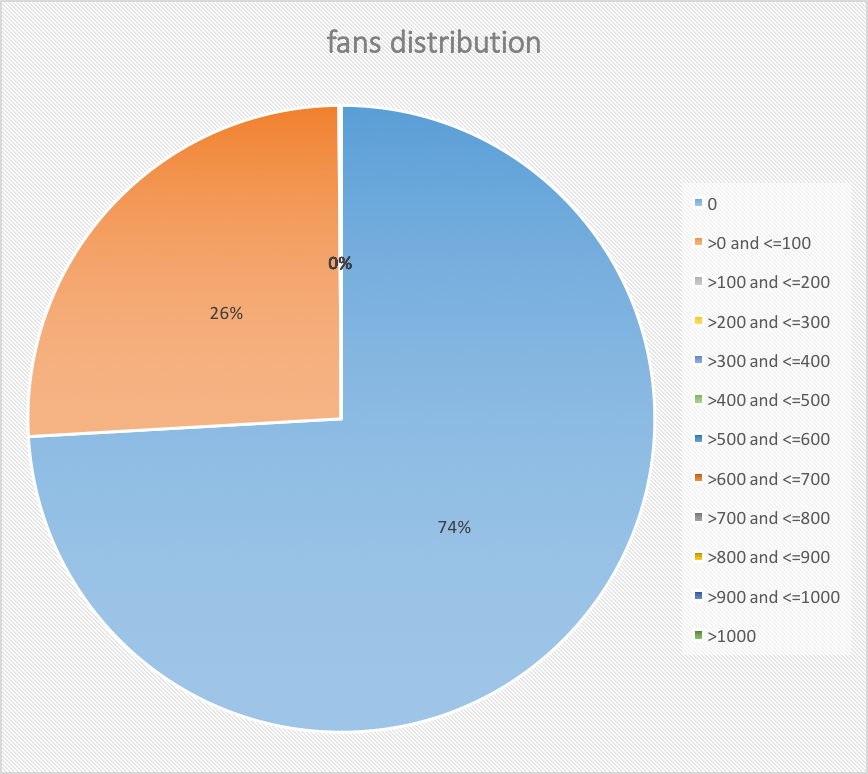}
        \caption{Fans Distribution}
        \label{fig:fans}
  \end{subfigure}
  \begin{subfigure}[b]{0.45\textwidth}        
        \includegraphics[width=0.99\textwidth]{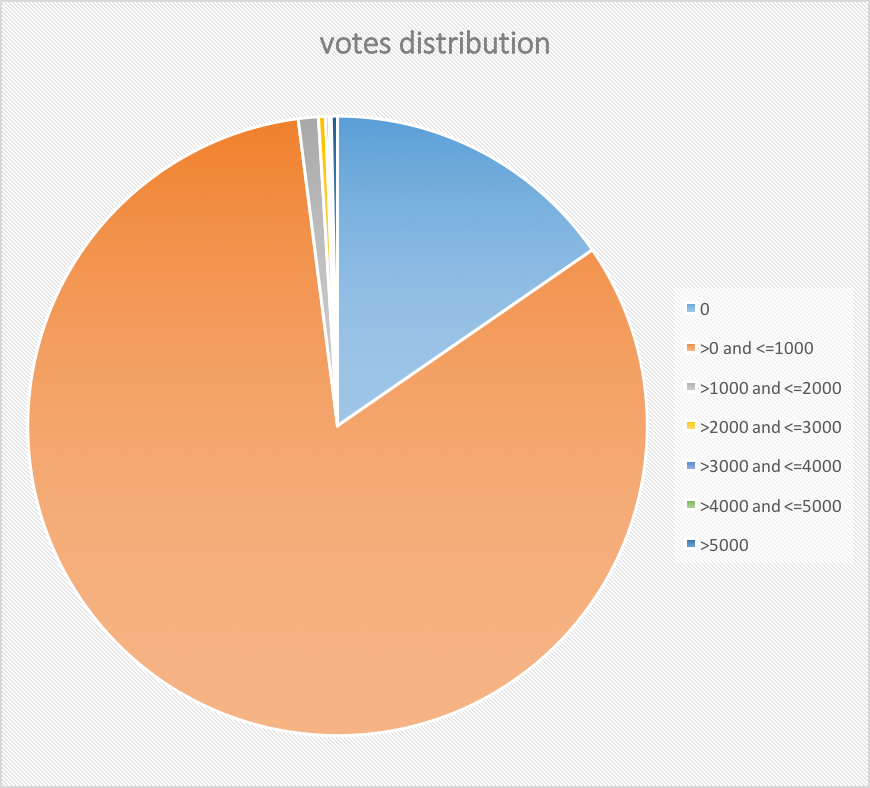}
        \caption{Votes Distribution}
        \label{fig:votes}
  \end{subfigure}  
\label{fig:dis}
\end{figure*}

\subsection{User Distribution in Stars, Reviews, Fans and Votes}
The distribution of users in terms of four important attributes (stars, reviews, fans and votes) 
are evaluated in this subsection. Before going into the details of the results, we introduce how these 
attributes are 
generated in Yelp.  Stars are used to evaluate the quality of a business after a visit. One can give stars from 1 to 5. 
The more the better; Reviews are comments from a user to a business, users usually write reviews to 
express their opinions; Fans are used when an user likes reviews or comments from another user 
(probability because they have similar taste)  and wants to receive more recommendations from him; 
Votes are comments to a review from a user. Specifically, user can submit a vote of \textit{useful} or 
\textit{cool} or \textit{funny} to a review.  

Figure~\ref{fig:star}-Figure~\ref{fig:votes} present the distributions of users for each attribute. The following details 
the analysis for each attribute. In terms of star distribution, businesses with 4 stars and 5 stars are the most
(41\% and 38\% respectively), which means that businesses collaborated with Yelp are high quality and users
have positive feedbacks to most of these businesses. Note that the results are highly consistent with the 
results released in the official website~\cite{yelpdata}. 

In terms of review distribution, besides the percentage of users who do not write any reviews, the percentage of 
users who write a particular number range of reviews (in the grain of 100) are also presented.  As indicated in 
Figure~\ref{fig:review}, the percentage of users who do not write reviews are negligible, although 53\% of all 
users do not have any friends on Yelp, which implies that review is a well-regarded feature on Yelp.  
If we look at the distribution of users who write reviews, it is easy to find that users who write no more than 
100 reviews dominate (92\%). This number tells us that although most of Yelp user do write reviews, 
most of them do not write a lot. 

Figure~\ref{fig:fans} presents the results for fans. The interesting finding is that most of Yelp users (74\%) 
do not have any fans, while users with 0 to 100 fans dominate the remaining percentage. From the results, 
we know that only a small portion of users actively comment a lot of businesses, which makes them followed 
by others. Most of users only care about the businesses that they have ever been experienced, which are limited.
 
Figure~\ref{fig:votes} shows the user breakdown in terms of votes.  Similar to Figure~\ref{fig:review}, users who 
vote 0 to 1000 are the largest part (83\%). The second largest part are users who do not vote (15\%). The 
total number of votes is also breakdown according to \textit{useful}, \textit{cool} and \textit{funny}. The percentage 
for each category is 48\%, 27\% and 25\%, respectively.  Among these three words, only \textit{funny} is 
negative in tone. Thus, most of the time(75\%), Yelp users satisfy with the quality of reviews. 

\subsection{Effectiveness of Elite User Mechanism}
A famous mechanism used to encourage review writing on Yelp is elite members~\cite{elite}. If one becomes
elite user successfully, he or she will be invited to many epic parties for free, which are not available for common 
users. Any Yelp user can apply for a elite user, but there is a  community to decide whether the application passes 
or not. Without doubt, elite users should be very active in many aspects such as review writing, commenting 
reviews from others and making friends on Yelp, although the internal algorithm to decide elite users are not
available to the public.  This subsection measures the effectiveness of elite user mechanism, which we believe 
important to Yelp because if elite users are not much more important than common users, Yelp should stop the 
investment to elite users as soon as possible to save cost.  

\begin{figure}[t]
  \centering
  \caption{The Percentage of Elite Users against Years} 
       \includegraphics[width=0.48\textwidth]{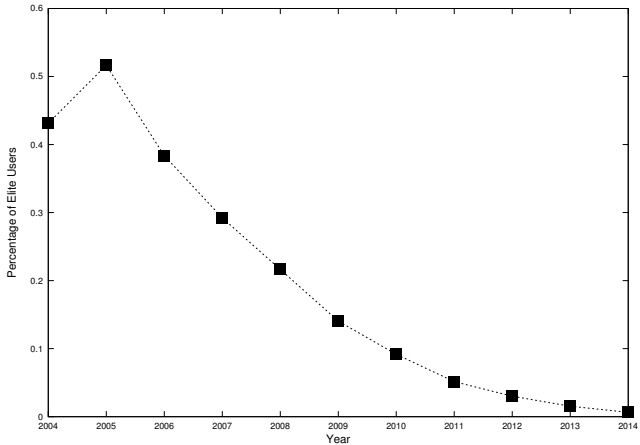}
  \label{fig:elite}
\end{figure} 
Before introducing the details of measurement, let us look at the percentage change of elite users against years.
Figure~\ref{fig:elite} presents the experimental results. As shown in the figure, there is an increase of the percentage 
in 2005, and the curve keeps decreasing since then. At 2005, the percentage is more than 50\%, while it is 
even smaller than 1\% in 2014. Note that Yelp is established in 2004. The curve suggests that at early stage of 
Yelp, the company uses this mechanism to engage users. Once the reputation is established, the increase of 
elite users is much slower than the increase of common users. 

Next, we start to measure the effectiveness of elite user mechanism. Our methodology is to collect several 
metrics of elite users and compare them to average users.
The first metric is the average number of fans for elite users, which is calculated to be 16. For all users, the average 
number of fans is 1.6, which is significantly smaller than 16. Next, we look at the average number of reviews written. 
The experimental results for elite users and all users are 245 and 32, respectively. Again, the metric for elite users 
is much better than the average of all users. The third metric we used is the average number of votes. The number is
1336 for elite users and 122 for all users. Finally, we compare the average degree for elite users. The result is 55, which 
is much larger than the average(7). From these four metrics, one can find elite users own significantly more fans 
(10$\times$), write much more reviews (7.6$\times$), comments more than others(10.9$\times$), and has more friends
(7.8$\times$). In another word, elite users are much more active in the community. Next, we try to answer an interesting 
question. Do elite users give more or less stars for a business, compared to common users? 
Similar to metrics collected, the average of stars given by 
elite users is 3.78, while the average is 3.72, when considering all users as a whole. Although the value for elite users
is higher, it is rather close to the average of all users. This result implies that elite users are famous for their activeness 
in the community(like the quality of reviews), instead of giving stars which are out of exceptions. Another insight of the 
measurement to elite users is that Yelp community can develop a machine learning model by the use of our metrics, 
which helps to decide whether an application for elite user passes or not, instead of evaluating it manually. 

\section{Related Work}
During the several rounds of Yelp challenge, a lot of interesting works are done to make use of the dataset. 
For instance, Bakhshi \textit{et al.} use the dataset to understand the difference of review feedbacks including 
\textit{useful}, \textit{cool} and \textit{funny}~\cite{cscw}. Lee \textit{et al.} targets for a multidimensional model, 
which are used to find expert (elite users) on Yelp~\cite{expert}. The model uses review text, network structure, 
and user metadata to find elite users. Their conclusion was the model needs to consider all information together 
to improve the accuracy of the model. In this paper, we also find several efficient indicators for elite users, 
which are complementary to their work. Crain~\textit{et al.} analyzed the elite users in terms of the network 
structure, where they found elite users are \textit{super-connectors}  in Yelp~\cite{stanford}. 
The results are consistent with ours. Besides, a lot more analysis are provided in our work. Tanvi Jindal proposed 
to find local experts by the use of machine learning~\cite{local}, while our focus in this paper is to analyze the 
dataset, find useful insights and be helpful to the design of next generation online review system.

\section{Conclusion}
This paper presents analysis results for Yelp dataset. Our main focus is user related data, such as the social network 
formed by users and attributes in user nodes.  We did not investigate the user behaviors of visiting a business and Yelp 
reviews, which are believed to be valuable.  These parts are left in future work. 

\bibliographystyle{plain}
\bibliography{ref} 

\end{document}